\documentclass{an}
\usepackage{graphicx}
\usepackage{times}
\Volume{01}              
\Year{2006}              
\Month{06}               
\Pagespan{1}{7}      	 

\begin{document}
\lhead[\thepage]{A.N. Bilir et al.: Separation of dwarf and giant stars with ROTSE-IIId}
\rhead[Astron. Nachr./AN~{\bf XXX} (200X) X]{\thepage}
\headnote{Astron. Nachr./AN {\bf 32X} (200X) X, XXX--XXX}

\title{Separation of dwarf and giant stars with ROTSE--IIId}

\author{S. Bilir
\and  T. G\"uver
\and  M. Aslan
}
\institute{Istanbul University Science Faculty, 
           Department of Astronomy and Space Sciences, 
           34119, University-Istanbul, Turkey}
\date{} 

\abstract{136 stars which were known to be the members of open cluster NGC 752 
were observed at $R$ band with ROTSE--IIId telescope located at the Turkish 
National Observatory (TUG) site. The data had been evaluated together with BV 
and 2MASS photometric data. A new practical method for separating dwarf and 
giant was described and applied. Evaluating the colour magnitude--diagrams with 
Padova isochrones revealed metallicity similar to the Sun and an age of 1.41 Gyr 
for the open cluster NGC 752.    
}
\correspondence{sbilir@istanbul.edu.tr}

\maketitle

\section{Introduction}

One of the problems of the Galactic astronomy is the estimation of Galactic model 
parameters of giant stars in our Galaxy. In many studies, the Galactic model 
parameters are estimated without any discrimination between dwarfs and giants, 
whereas some researchers estimated model parameters only for certain star categories 
(e.g. Pritchet 1983, Bahcall \& Soneira 1984, Buser \& Kaeser 1985 and Mendez \& 
Altena 1996). A very recent work is an example for this where the Galactic model 
parameters were estimated using only giants (Cabrera-Lavers, Garzon \& Hammersley 
2005). The separation of field dwarfs and field giants plays an important role for such 
kinds of works. The most efficient classical methods of identifying dwarf and giant 
stars utilize spectroscopy. By inspecting spectral line profiles, one has to 
estimate surface gravity to discriminate between higher and lower pressure stellar 
atmospheres to be sure for the identification. This is, however time consuming and 
tiring. A rather easier procedure is to separate dwarfs and evolved stars (subgiants 
or giants) such as to obtain a luminosity function consistent with the local 
luminosity function of nearby stars due to Gliese \& Jahreiss (1991) and Jahreiss 
\& Wielen (1997). The procedure of this separation is based on the fact that the 
local luminosity functions obtained for many fields indicates a systematic excess 
of star counts relative to the luminosity function of nearby stars for the fainter 
segment, i.e. $M(V)\geq5^{m}.5$, and a deficit for brighter segment, $M(V)<5^{m}.5$. 
(in $RGU$ system $M(G)\geq6^{m}$ and $M(G)<6^{m}$, respectively). The works of 
Karaali (1992); Ak, Karaali \& Buser (1998); Karata\c{s}, Bilir \& Karaali 
(2000); Karaali et al. (2000); Karata\c{s}, Karaali \& Buser (2001); Karaali, 
Bilir \& Buser (2004); Bilir, Karaali \& Buser (2004) and Karata\c{s} et al. 
(2004) can be given as examples for application of this procedure.  

Recently, a new method were suggested by Bilir et al. (2006) for separating the 
field dwarfs and field giants. This new method is based on the comparison of the Two 
Micron All Sky Survey (2MASS, hereafter) $J$, $H$, $K_{s}$ with the $V$ magnitudes 
down to the limiting magnitude of $V=16$. In this work we extend application of 
this method to the open cluster NGC 752 observed by a robotic telescope ROTSE-IIId 
(Akerlof et al. 2003).

This paper is organized as follows. In Section 2 the BV, 2MASS and ROTSE data 
are presented. In Section 3 the method is applied to ROTSE and 2MASS data, and 
the separation of dwarf and giant stars is tested. In Section 4 colour-magnitude 
diagrams (CMDs) of NGC 752 is compared to the Padova isochrones. Finally, the 
conclusion is given in Section 5. 

\begin{figure}
\center
\resizebox{5cm}{7cm}{\includegraphics*{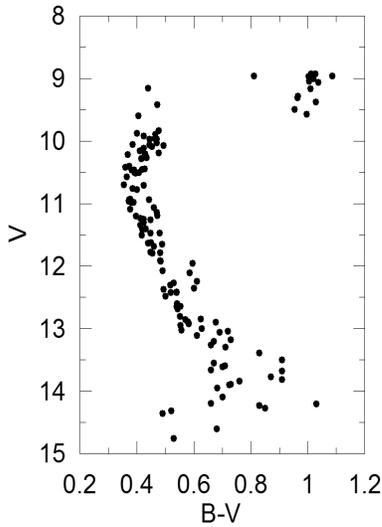}} 
\caption {$V \times (B-V)$ CMD of 136 probable member stars in NGC 752.}
\end {figure}

\begin{figure*}
\center
\resizebox{17cm}{5.51cm}{\includegraphics*{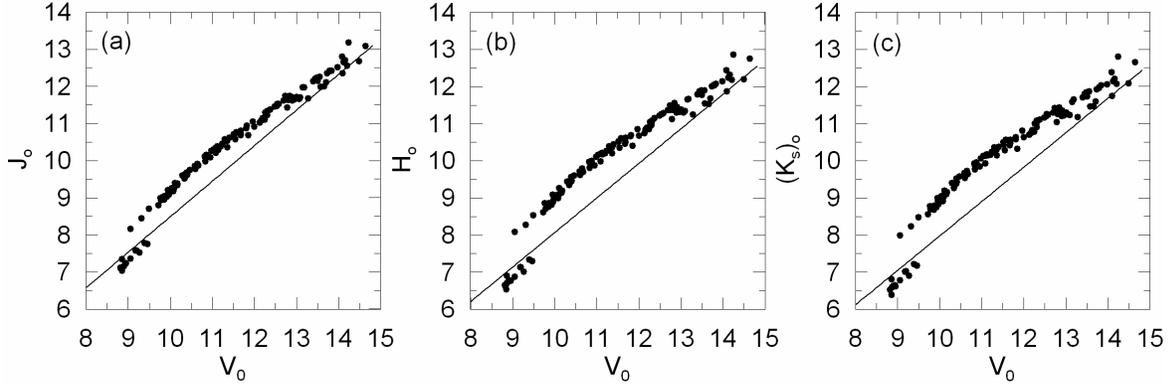}} 
\caption {V and 2MASS magnitudes of 136 stars in NGC 752. (a) $J_{0} \times V_{0}$, 
(b) $H_{0} \times V_{0}$, and (c) $Ks_{0} \times V_{0}$. The solid lines in diagrams 
were drawn according to eqs. (1), (2), and (3).}
\end {figure*}

\section{Observations}
\subsection{The BV and 2MASS Data}

NGC 752=C0154+374 ($\alpha=01^{h}57^{m}41^{s}$, $\delta=+37^{o}47^{'}06^{''}$; 
$l=137^{o}.13$, $b=-23^{o}.25$; J2000) has been subject of many studies, because 
it is the nearest intermediate-age cluster, with 427 pc (Dzervitis \& Paupers 
1993) distance from the Sun. It is usually considered as metal-deficient with respect 
to the Sun, $[Fe/H]=-0.15 \pm$ 0.05 dex, slightly reddened $E(B-V)=0.035 \pm 0.005$, 
with distance module $(m-M)=8.25\pm$0.10 (Daniel et al. 1994). Accurate proper motion 
and radial velocity measurements show that there are 136 probable member stars of the 
open cluster (Daniel et al. 1994). $V$ magnitudes and $(B-V)$ colour indices used in 
this study were taken from Daniel et al. (1994) and they were given in Table 1. The 
$V\times(B-V)$ CMD in Fig. 1 shows that stars with $V<10$ and $(B-V)>0.80$ are giants.

Recently the 2MASS, including the Point-Source Catalogue and Atlas, has produced huge 
amounts of data to be explored in the coming years (Skrutskie et al. 1997). The 
photometric system comprises Johnson's $J$ (1.25 $\mu$m) and $H$ (1.65 $\mu$m) bands with 
the addition of $K_{s}$ (2.17 $\mu$m), slightly bluer than Johnson's K. The 2MASS sky 
coverage, homogeneity and depth will certainly make this set of filters a photometric 
standard reference for the future.

2MASS data of the 136 probable member stars in NGC 752 are obtained by 
Vizier\footnote{http://vizier.u-strasbg.fr/viz-bin/Vizier?-source=2MASS} in CDS 
and they are given in Table 1. We used the equations of Fiorucci \& Munari (2003) 
for the determination of the total absorptions for the bands $V$, $J$, $H$ and 
$K_{s}$, i.e. $A(V)=3.1E(B-V)$, $A(J)=0.887E(B-V)$, $A(H)=0.565E(B-V)$ and 
$A(K_{s})=0.382E(B-V)$. Thus the de-reddened magnitudes were obtained as follows:  
$V_{0}=V-A(V)$, $J_{0}=J-A(J)$, $H_{0}=H-A(H)$ and $(K_{s})_{0}=K_{s}-A(K_{s})$.
The subscript ``0'' indicates de-reddened magnitude.

$J_{0} \times V_{0}$, $H_{0} \times V_{0}$ and $(K_{s})_{0} \times V_{0}$ diagrams 
for the cluster stars are given in Fig. 2. The solid lines represent the equations 
in Bilir et al. (2006), i.e. 

\begin{equation}
J_{0} = 0.957V_{0} - 1.079,
\end{equation}

\begin{equation}
H_{0} = 0.931V_{0} - 1.240,
\end{equation}

\begin{equation}
(K_{s})_{0} = 0.927V_{0} - 1.292.	
\end{equation}

14 stars below the lines are the giants in Fig. 1, whereas 122 stars above the 
lines are the dwarfs of the same cluster. The distribution of different star 
categories at different sides of the lines in three figures were presented here 
to confirm separation of dwarfs and giants. 

\subsection{ROTSE data}

The Robotic Optical Transient Experiment (ROTSE-III) consists of four 0.45m 
worldwide robotic, automated telescopes situated at different locations on Earth. 
They are designed for fast ($\sim$ 6 sec) responses to Gamma-Ray Burst (GRB) 
triggers from satellites such as Swift. Each ROTSE telescope has a $1.85 \times 1.85$ 
deg$^{2}$ field of view, and uses a Marconi 2048 $\times$ 2048 back illuminated 
thinned CCD. These telescopes operate without filters, and have wide passband which 
peaks around 550 nm (Akerlof et al. 2003). In this work, we present optical 
observations of NGC 752 performed by ROTSE-IIId, telescope located at Turkish National 
Observatory (TUG) site, Bak$\i$rl$\i$tepe, Turkey. The observations took place between 
MJD 53637 (September 2005) and MJD 53649 (October 2005). A total of about 217 CCD 
frames were analyzed. After determining the instrumental magnitudes (Bertin \& Arnouts, 
1996), they were reduced to ROTSE magnitudes via comparing all the field stars with 
the USNO A2.0 $R$-band catalog. All the processes were done in an automated mode. 

\begin{figure}
\center
\resizebox{6.51cm}{6.22cm}{\includegraphics*{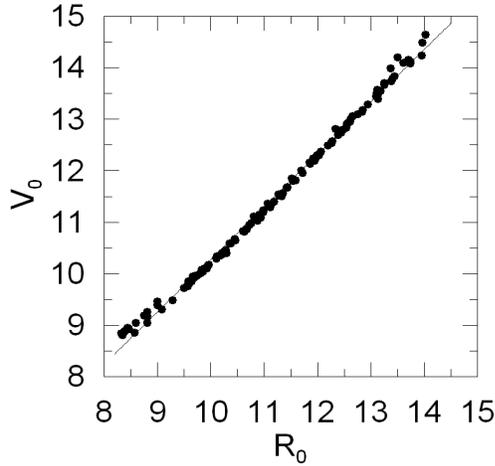}} 
\caption {$R_{0} \times V_{0}$ magnitudes of 136 stars in NGC 752.}
\end {figure}

\begin{figure*}
\center
\resizebox{17cm}{5.51cm}{\includegraphics*{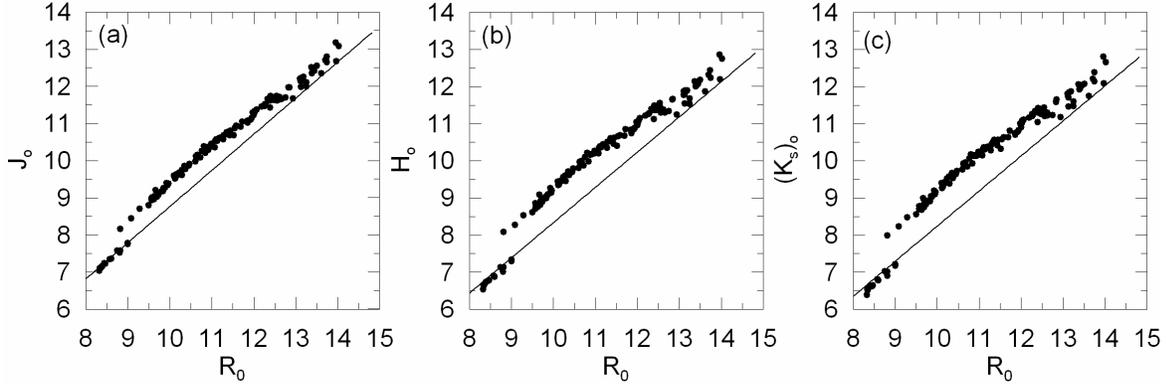}} 
\caption {$R_{0}$ and 2MASS magnitudes of 136 stars in NGC 752. (a) $J_{0} \times R_{0}$, 
(b) $H_{0} \times R_{0}$, and (c) $Ks_{0} \times R_{0}$. The solid lines in diagrams were 
drawn according to eqs. (5), (6), and (7).}
\end {figure*}

$R$-band magnitudes of the 136 stars are given in Table 1. ROTSE magnitudes are also 
de-reddened in order to homogenize the data. The total absorption for the $R$ band 
could be determined by the equation of Fiorucci \& Munari (2003), i.e. 
$A(R)=2.613E(B-V)$. Thus the de-reddened magnitude in $R$ becomes $R_{0}=R-A(R)$.  

\section{Application of the method to ROTSE-IIId data}
We used the following relation between the $V_{0}$ magnitude and the ROTSE-IIId 
magnitude $R_{0}$ for 136 stars of the open cluster NGC 752 (Fig. 3) in order to 
apply the method to ROTSE-IIId data:

\begin{equation}
V_{0}=(1.019\pm0.006)R_{0}+0.092\pm0.070~~(\sigma=0.10).	
\end{equation}
Thus, substituting the value of $V_{0}$ in (4) into (1), (2), and (3) we obtain the 
relations between 2MASS and ROTSE-IIId magnitudes, i.e.   

\begin{equation}
J_{0} = 0.975R_{0} - 0.991,
\end{equation}

\begin{equation}
H_{0} = 0.949R_{0} - 1.154,
\end{equation}

\begin{equation}
(K_{s})_{0}= 0.945R_{0} - 1.207.	
\end{equation}
The diagrams $J_{0} \times R_{0}$, $H_{0} \times R_{0}$ and $(K_{s})_{0} \times R_{0}$ 
for the NGC 752 cluster stars and the line corresponding to the eqs. (5), (6), and (7) 
are given in Fig. 4. One can see that dwarfs and giants lie at opposite sides of the 
line in these figures, especially the relation between $(K_{s})_{0} \times R_{0}$ 
(Fig. 4c) is the most successful in separating of the two different star categories. 
 
\section{Age estimation for the open cluster NGC 752 via two photometries}

We estimated the age of the open cluster NGC 752 by the means of BV and 2MASS 
data to show the advantage. The absolute magnitudes of the stars were determined 
by the corresponding apparent magnitudes and the distance module of the cluster. 
The Padova isochrones were taken from Girardi et al. (2002) and Bonatto, Bica \& 
Girardi (2004) for BV\footnote
{http://pleiadi.pd.astro.it/isoc\_photsys.02/isoc\_photsys.02.html} and 2MASS\footnote{
http://pleiadi.pd.astro.it/isoc\_photsys.01/isoc\_2mass/index.html} 
photometry, respectively. The isochrone sets were computed with updated opacities, 
and equations of state, and a moderate amount of convective overshoot. The basic 
isochrone set presented in Girardi et al. (2002) covers a very wide range of initial 
masses (from 0.15 to $\sim 100 M_{\odot}$), metallicities, and photometric systems, 
being well suited for studies of clusters of all ages.
 
The isochrones mentioned above were fitted to the CMDs in Figs. 5-7 for two sets 
of chemical compositions, i.e. Z=0.019, Y=0.273 (panel a) and Z=0.008, Y=0.250 
(panel b). The isochrones in Fig. 5a fits to the main-sequence and turn-off segments 
of the $M_{V} \times (B-V)_{0}$ diagram and reveal an age of $t=1.26$ Gyr, whereas 
in Fig. 5b, the isochrones could be fitted only to the giant branch of the same CMD, 
resulting a larger age, i.e. $t=1.78$ Gyr. The isochrones could not be fitted to 
all segments of the $M_{J} \times (J-H)_{0}$ diagram in Fig. 6 either. In Fig. 6a 
the fit is better to the main-sequence and turn-off segments, however it is only to 
the giant branch in Fig. 6b. The best fit is accomplished with the isochrone of age 
$t=1.41$ Gyr to the $M_{J} \times (J-K_{s})_{0}$ in Fig. 7a. In fact, the isochrone 
fits to all segments, main-sequence, turn-off and giant branch, for a metallicity 
close to the solar one which is expected (Daniel et al. 1994). Thus, the comparison 
of the six diagrams in Figs. 5-7 reveals that the CMD $M_{J} \times (J-K_{s})_{0}$ 
is the best one which fits for the age estimation. 

\begin{figure}
\center
\resizebox{8cm}{6.92cm}{\includegraphics*{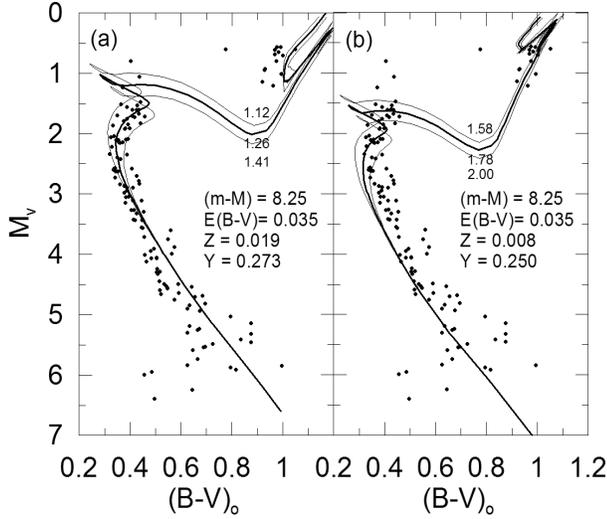}} 
\caption {$M_{V} \times (B-V)_{0}$ CMDs of NGC 752. (a) Z=0.019 and 1.12, 1.26 and 1.41 
Gyr and (b) Z=0.008 and 1.58, 1.78 and 2.00 Gyr of Padova isochrones.}
\end {figure}

\begin{figure}
\center
\resizebox{8cm}{6.92cm}{\includegraphics*{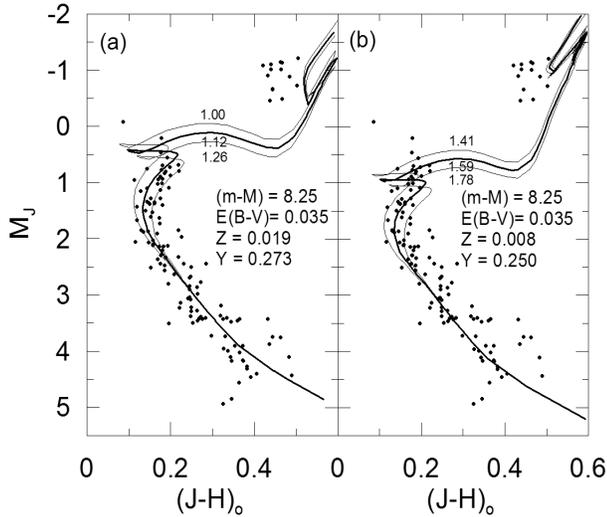}} 
\caption {$M_{J} \times (J-H)_{0}$ CMDs of NGC 752. (a) Z=0.019 and 1.00, 1.12 and 1.26 
Gyr and (b) Z=0.008 and 1.41, 1.59 and 1.78 Gyr of Padova isochrones.}
\end {figure}

\begin{figure}
\center
\resizebox{8cm}{6.92cm}{\includegraphics*{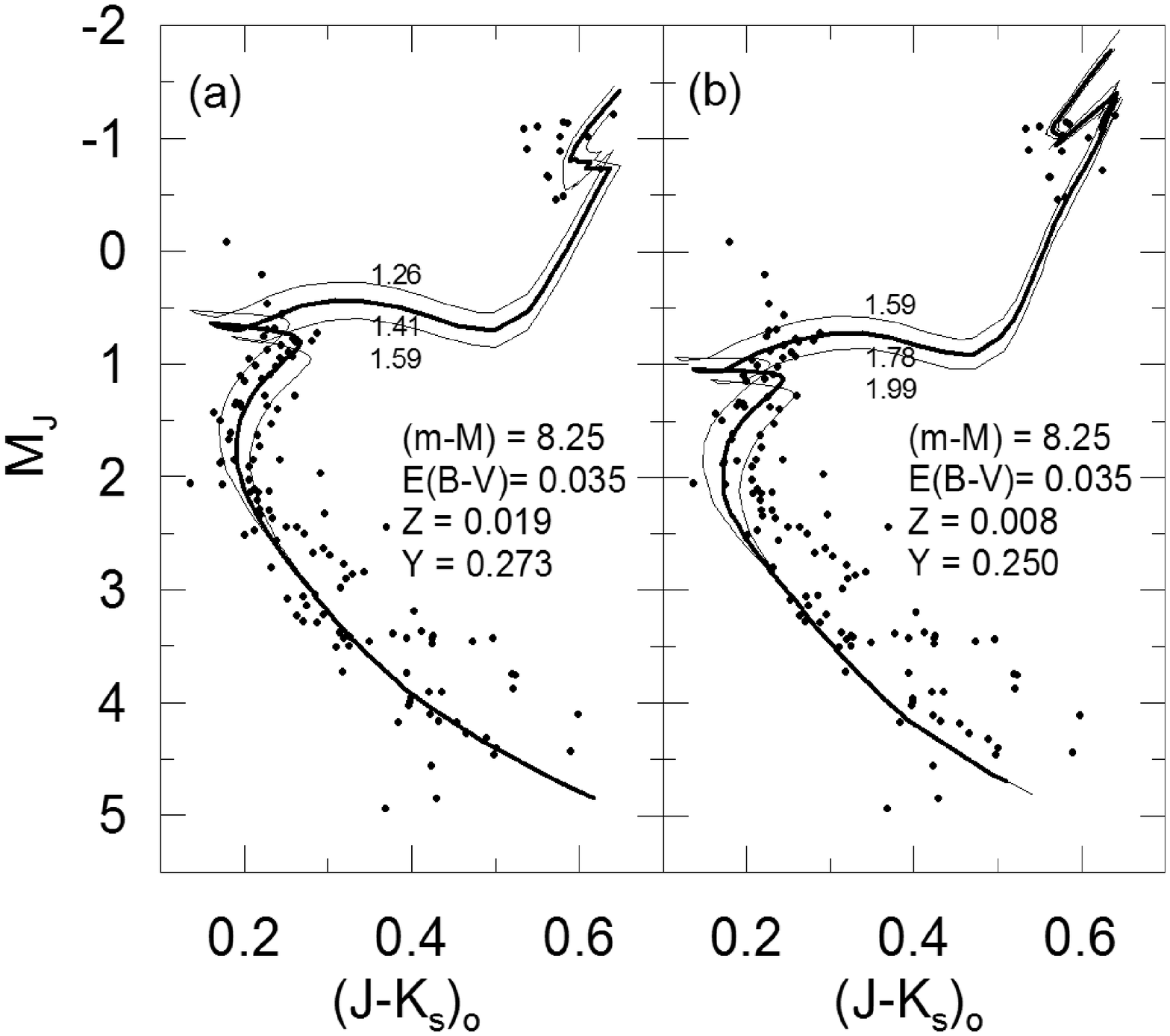}} 
\caption {$M_{J} \times (J-Ks)_{0}$ CMDs of NGC 752. (a) Z=0.019 and 1.26, 1.41 and 1.59 
Gyr and (b) Z=0.008 and 1.59, 1.78 and 1.99 Gyr of Padova isochrones.}
\end {figure}

\section{Conclusion}
In this study, we have reduced the relations between the 2MASS and $V$ magnitudes by 
which field dwarfs and giants can be separated, to the USNO A2.0 $R$-band magnitudes. 
The $R_{0}$ magnitudes of 136 stars in open cluster NGC 752 were transferred to 
the $V_{0}$ magnitudes, and the relations between $J_{0}$, $H_{0}$, $(K_{s})_{0}$ 
and $R_{0}$ were derived by means of the relations between $J_{0}$, $H_{0}$,  
$(K_{s})_{0}$ and $V_{0}$ given by Bilir et al. (2006). Dwarf and giant stars 
identified by the CMD of open cluster NGC 752 lie at different sides of the line 
representing the relation between the 2MASS and $R_{0}$ magnitudes, in the 
$J_{0} \times R_{0}$, $H_{0} \times R_{0}$ and $(K_{s})_{0} \times R_{0}$ 
diagrams. The best one is the last diagram, i.e. $(K_{s})_{0} \times R_{0}$. 
Thus, dwarf-giant separation could be carried out also in the ROTSE-IIId data. 
Proven to be successful, this practical method can provide good contributions to 
the studies of Galactic model parameters in which separation of dwarfs and giants 
were needed.  

A set of Padova isochrones were fitted to the CMDs of the open cluster NGC 
752 using BV and 2MASS photometric data. It turned out that the isochrone with 
chemical composition Z=0.019 and Y=0.273 which reveals an age of 1.41 Gyr 
for the open cluster NGC 752 could be fitted to all segments, i.e. 
main-sequence, turn-off and giant branch, of the $M_{J} \times (J-K_{s})_{0}$ 
two-colour diagram. This result is very close to the age 1.24$\pm$0.20 
Gyr which Salaris, Weiss \& Percival (2004) calculated from the morphology 
of 71 open clusters in our Galaxy. The metal-abundance of the cluster given 
by Daniel et al. (1994), i.e. $[Fe/H]=-0.15\pm0.05$ dex, is a strong 
confirmation for our result.

\begin{acknowledgements}
We thank international ROTSE collaboration and TUG for the optical facilities 
(Project number: TUG-ROTSE.05.14). This research has made use of the SIMBAD 
database, operated at CDS, Strasbourg, France. This publication makes use of 
data products from the Two Micron All Sky Survey, which is a joint project of 
the University of Massachusetts and the Infrared Processing and Analysis 
Center/California Institute of Technology, funded by the National Aeronautics 
and Space Administration and the National Science Foundation. We would also 
like to thank Dr. Salih Karaali for helpful comments and suggestions, Dr. Zeki 
Eker for reading the whole manuscript and correction and Dr. Tansel Ak for 
various helps. This work was supported by the Research Fund of the University 
of Istanbul. Project number: BYP 914.

\end{acknowledgements}

\begin{table*}
{\scriptsize
\center
\caption{BV, ROTSE and 2MASS magnitudes and its errors of 136 probable member stars of open cluster NGC 752. ID name is the same as Daniel et al. (1994) and the coordinates are for the epoch 2000.}  
\begin{tabular}{rcccccccccccccc}
\hline
 ID & $\alpha$ & $\delta$ & \multicolumn{2} {c} {V} & \multicolumn{2} {c} {B-V}& \multicolumn{2} {c} {R} & \multicolumn{2} {c} {J}& \multicolumn{2} {c} {H}& \multicolumn{2} {c} {$K_{s}$}\\
\hline
 143 & 01 54 31.04 & +37 29 31.6 & 11.140 &       & 0.470 &       & 10.970 & 0.010 & 10.343 & 0.019 & 10.184 & 0.024 & 10.151 & 0.020 \\
 215 & 01 54 56.71 & +37 58 28.9 & 11.400 &       & 0.430 &       & 11.200 & 0.012 & 10.615 & 0.021 & 10.416 & 0.017 & 10.378 & 0.018 \\
 222 & 01 54 59.65 & +37 28 59.8 & 11.651 & 0.009 & 0.489 & 0.007 & 11.365 & 0.011 & 10.607 & 0.021 & 10.377 & 0.018 & 10.293 & 0.021 \\
 237 & 01 55 02.85 & +38 16 38.1 & 12.418 & 0.000 & 0.519 & 0.000 & 12.110 & 0.017 & 11.364 & 0.021 & 11.105 & 0.017 & 11.094 & 0.020 \\
 245 & 01 55 07.13 & +37 32 36.9 & 14.190 &       & 0.660 &       & 13.830 & 0.026 & 12.838 & 0.022 & 12.464 & 0.021 & 12.397 & 0.021 \\
 259 & 01 55 12.62 & +37 50 14.4 &  9.496 & 0.013 & 0.954 & 0.006 &  9.092 & 0.016 &  7.818 & 0.023 &  7.370 & 0.023 &  7.228 & 0.016 \\
 264 & 01 55 15.29 & +37 50 31.2 &  9.569 & 0.005 & 0.996 & 0.003 &  9.090 & 0.017 &  7.795 & 0.023 &  7.317 & 0.016 &  7.197 & 0.018 \\
 300 & 01 55 26.18 & +38 08 22.0 & 13.610 &       & 0.700 &       & 13.200 & 0.019 & 12.182 & 0.018 & 11.798 & 0.021 & 11.743 & 0.021 \\
 305 & 01 55 27.68 & +37 34 04.5 & 10.152 & 0.003 & 0.411 & 0.001 &  9.940 & 0.011 &  9.293 & 0.018 &  9.103 & 0.015 &  9.062 & 0.016 \\
 308 & 01 55 27.67 & +37 59 55.2 &  9.285 & 0.011 & 0.966 & 0.009 &  8.902 & 0.020 &  7.618 & 0.021 &  7.163 & 0.021 &  7.039 & 0.020 \\
 313 & 01 55 29.29 & +37 50 26.2 &  9.968 & 0.013 & 0.470 & 0.001 &  9.718 & 0.014 &  9.001 & 0.023 &  8.792 & 0.018 &  8.696 & 0.017 \\
 350 & 01 55 39.37 & +37 52 52.4 &  8.922 & 0.002 & 1.011 & 0.006 &  8.439 & 0.024 &  7.151 & 0.020 &  6.674 & 0.016 &  6.547 & 0.016 \\
 356 & 01 55 42.39 & +37 37 54.3 &  9.161 & 0.003 & 1.010 & 0.001 &  8.687 & 0.020 &  7.395 & 0.019 &  6.900 & 0.020 &  6.800 & 0.018 \\
 361 & 01 55 44.75 & +37 54 42.5 & 13.899 & 0.000 & 0.724 & 0.000 & 13.497 & 0.025 & 12.438 & 0.023 & 12.054 & 0.023 & 11.988 & 0.019 \\
 363 & 01 55 44.93 & +38 08 21.4 & 10.420 &       & 0.360 &       & 10.189 & 0.013 &  9.619 & 0.021 &  9.464 & 0.021 &  9.410 & 0.020 \\
 372 & 01 55 47.40 & +37 42 26.4 &  9.890 & 0.005 & 0.464 & 0.000 &  9.656 & 0.011 &  8.971 & 0.029 &  8.741 & 0.027 &  8.717 & 0.018 \\
 391 & 01 55 53.54 & +37 49 26.6 & 13.890 &       & 0.730 &       & 13.501 & 0.024 & 12.457 & 0.023 & 12.072 & 0.021 & 11.985 & 0.023 \\
 397 & 01 55 55.53 & +37 28 33.2 &  9.831 & 0.000 & 0.477 & 0.005 &  9.585 & 0.011 &  8.839 & 0.021 &  8.636 & 0.020 &  8.576 & 0.016 \\
 413 & 01 55 59.44 & +37 40 48.5 & 12.303 & 0.016 & 0.517 & 0.022 & 12.035 & 0.013 & 11.181 & 0.021 & 10.924 & 0.017 & 10.842 & 0.016 \\
 429 & 01 56 02.95 & +37 36 32.7 & 14.270 &       & 0.850 &       & 13.791 & 0.033 & 12.741 & 0.023 & 12.338 & 0.022 & 12.225 & 0.021 \\
 435 & 01 56 03.69 & +37 59 22.4 & 11.467 & 0.017 & 0.447 & 0.010 & 11.223 & 0.013 & 10.572 & 0.020 & 10.384 & 0.018 & 10.324 & 0.018 \\
 455 & 01 56 08.96 & +37 39 52.6 & 10.512 & 0.013 & 0.395 & 0.003 & 10.298 & 0.020 &  9.657 & 0.021 &  9.496 & 0.017 &  9.442 & 0.014 \\
 461 & 01 56 10.30 & +37 44 59.9 & 10.054 & 0.002 & 0.384 & 0.006 &  9.752 & 0.041 &  9.239 & 0.021 &  9.115 & 0.018 &  9.015 & 0.019 \\
 465 & 01 56 11.11 & +37 45 11.1 & 11.229 & 0.012 & 0.412 & 0.007 & 10.896 & 0.045 & 10.419 & 0.021 & 10.235 & 0.019 & 10.194 & 0.018 \\
 472 & 01 56 12.88 & +38 01 43.2 & 11.060 &       & 0.460 &       & 10.813 & 0.140 & 10.127 & 0.020 &  9.897 & 0.018 &  9.866 & 0.020 \\
 475 & 01 56 13.70 & +37 15 56.9 & 12.847 & 0.000 & 0.624 & 0.000 & 12.496 & 0.016 & 11.778 & 0.022 & 11.514 & 0.026 & 11.435 & 0.022 \\
 477 & 01 56 13.96 & +37 47 04.7 & 10.572 & 0.006 & 0.364 & 0.006 & 10.375 & 0.015 &  9.778 & 0.020 &  9.627 & 0.019 &  9.589 & 0.021 \\
 479 & 01 56 14.28 & +37 58 14.2 & 10.938 & 0.011 & 0.442 & 0.005 & 10.696 & 0.014 & 10.013 & 0.021 &  9.828 & 0.018 &  9.777 & 0.017 \\
 486 & 01 56 15.51 & +37 38 41.5 & 10.074 & 0.000 & 0.493 & 0.006 &  9.824 & 0.012 &  9.115 & 0.026 &  8.918 & 0.017 &  8.853 & 0.016 \\
 505 & 01 56 18.63 & +37 37 39.3 & 10.778 & 0.005 & 0.399 & 0.008 & 10.543 & 0.014 &  9.903 & 0.021 &  9.719 & 0.017 &  9.669 & 0.016 \\
 506 & 01 56 18.90 & +37 58 00.4 &  8.971 & 0.018 & 1.003 & 0.011 &  8.456 & 0.044 &  7.174 & 0.018 &  6.723 & 0.024 &  6.606 & 0.023 \\
 512 & 01 56 21.65 & +37 36 08.2 &  9.375 & 0.022 & 1.029 & 0.004 &  8.896 & 0.018 &  7.558 & 0.020 &  7.045 & 0.020 &  6.915 & 0.018 \\
 517 & 01 56 22.57 & +37 39 17.8 & 14.230 &       & 0.830 &       & 13.833 & 0.030 & 12.681 & 0.021 & 12.265 & 0.024 & 12.163 & 0.022 \\
 520 & 01 56 23.10 & +37 38 03.0 & 12.850 &       & 0.570 &       & 12.530 & 0.016 & 11.684 & 0.021 & 11.390 & 0.023 & 11.342 & 0.022 \\
 542 & 01 56 29.45 & +37 55 14.7 & 14.350 &       & 0.490 &       & 14.045 & 0.032 & 13.214 & 0.024 & 12.878 & 0.025 & 12.828 & 0.026 \\
 552 & 01 56 32.05 & +37 34 22.2 & 12.921 & 0.016 & 0.581 & 0.036 & 12.425 & 0.025 & 11.696 & 0.021 & 11.417 & 0.023 & 11.351 & 0.020 \\
 555 & 01 56 32.96 & +37 56 46.4 & 11.786 & 0.001 & 0.482 & 0.006 & 11.514 & 0.014 & 10.779 & 0.020 & 10.519 & 0.015 & 10.489 & 0.018 \\
 563 & 01 56 34.46 & +38 08 49.4 & 13.680 &       & 0.910 &       & 13.216 & 0.020 & 12.020 & 0.019 & 11.565 & 0.017 & 11.483 & 0.022 \\
 575 & 01 56 36.88 & +37 45 12.7 & 13.840 &       & 0.760 &       & 13.477 & 0.029 & 12.384 & 0.020 & 12.030 & 0.019 & 11.944 & 0.018 \\
 580 & 01 56 39.22 & +37 51 41.1 & 10.398 & 0.007 & 0.373 & 0.005 & 10.194 & 0.015 &  9.630 & 0.020 &  9.458 & 0.017 &  9.416 & 0.018 \\
 619 & 01 56 47.60 & +37 24 30.4 & 10.280 & 0.022 & 0.415 & 0.005 & 10.054 & 0.013 &  9.412 & 0.020 &  9.226 & 0.018 &  9.173 & 0.019 \\
 622 & 01 56 48.61 & +37 29 11.2 & 10.503 & 0.001 & 0.391 & 0.007 & 10.321 & 0.013 &  9.714 & 0.020 &  9.584 & 0.021 &  9.533 & 0.020 \\
 626 & 01 56 49.77 & +38 01 21.7 &  9.158 & 0.016 & 0.439 & 0.006 &  8.902 & 0.032 &  8.202 & 0.026 &  8.105 & 0.057 &  8.005 & 0.024 \\
 630 & 01 56 50.44 & +38 01 58.1 &  8.961 & 0.011 & 0.811 & 0.020 &  8.661 & 0.059 &  7.379 & 0.021 &  6.936 & 0.020 &  6.824 & 0.023 \\
 641 & 01 56 53.06 & +37 52 09.3 & 10.270 & 0.004 & 0.434 & 0.004 & 10.036 & 0.014 &  9.371 & 0.020 &  9.164 & 0.015 &  9.122 & 0.018 \\
 648 & 01 56 54.33 & +37 23 51.9 & 12.108 & 0.000 & 0.585 & 0.000 & 11.782 & 0.014 & 10.952 & 0.021 & 10.693 & 0.019 & 10.652 & 0.018 \\
 653 & 01 56 55.38 & +38 04 45.8 & 12.410 &       & 0.540 &       & 12.088 & 0.013 & 11.262 & 0.019 & 10.986 & 0.017 & 10.929 & 0.020 \\
 654 & 01 56 55.77 & +37 47 59.3 & 11.196 & 0.000 & 0.396 & 0.014 & 11.022 & 0.015 & 10.385 & 0.020 & 10.216 & 0.017 & 10.155 & 0.018 \\
 655 & 01 56 56.15 & +38 08 16.2 & 13.040 &       & 0.720 &       & 12.665 & 0.017 & 11.706 & 0.021 & 11.370 & 0.019 & 11.294 & 0.018 \\
 659 & 01 56 56.35 & +37 39 51.3 & 10.116 & 0.003 & 0.424 & 0.001 &  9.898 & 0.014 &  9.220 & 0.018 &  9.039 & 0.017 &  8.959 & 0.018 \\
 667 & 01 56 57.59 & +37 23 20.5 & 10.925 & 0.021 & 0.377 & 0.004 & 10.720 & 0.170 & 10.131 & 0.021 &  9.990 & 0.021 &  9.925 & 0.018 \\
 682 & 01 57 02.51 & +37 53 07.7 & 11.255 & 0.012 & 0.447 & 0.002 & 11.007 & 0.014 & 10.332 & 0.019 & 10.205 & 0.023 & 10.178 & 0.024 \\
 684 & 01 57 02.81 & +38 14 03.5 & 12.480 &       & 0.500 &       & 12.153 & 0.015 & 11.415 & 0.021 & 11.180 & 0.019 & 11.123 & 0.020 \\
 687 & 01 57 03.12 & +38 08 02.6 &  8.927 & 0.002 & 1.026 & 0.004 &  8.424 & 0.024 &  7.136 & 0.024 &  6.663 & 0.015 &  6.537 & 0.017 \\
 689 & 01 57 03.19 & +37 55 44.5 & 11.788 & 0.005 & 0.455 & 0.009 & 11.529 & 0.015 & 10.838 & 0.019 & 10.641 & 0.019 & 10.582 & 0.020 \\
 694 & 01 57 03.64 & +38 05 11.7 & 11.779 & 0.011 & 0.448 & 0.007 & 11.520 & 0.013 & 10.720 & 0.021 & 10.478 & 0.019 & 10.439 & 0.019 \\
 699 & 01 57 04.89 & +38 07 33.1 & 13.001 & 0.024 & 0.627 & 0.048 & 12.647 & 0.016 & 11.652 & 0.021 & 11.322 & 0.018 & 11.222 & 0.018 \\
 701 & 01 57 05.47 & +37 50 42.8 & 13.060 &       & 0.690 &       & 12.700 & 0.017 & 11.684 & 0.019 & 11.339 & 0.021 & 11.241 & 0.020 \\
 720 & 01 57 10.50 & +38 02 06.6 & 12.367 & 0.020 & 0.494 & 0.034 & 12.064 & 0.013 & 11.322 & 0.020 & 11.066 & 0.019 & 11.019 & 0.018 \\
 722 & 01 57 10.35 & +37 25 55.3 & 13.500 &       & 0.910 &       & 13.221 & 0.021 & 12.178 & 0.019 & 11.813 & 0.019 & 11.724 & 0.018 \\
 723 & 01 57 10.53 & +37 27 26.6 & 13.770 &       & 0.870 &       & 13.339 & 0.033 & 12.030 & 0.019 & 11.557 & 0.021 & 11.489 & 0.020 \\
 728 & 01 57 12.13 & +37 59 24.8 &  9.420 & 0.005 & 0.471 & 0.004 &  9.169 & 0.011 &  8.485 & 0.020 &  8.296 & 0.016 &  8.246 & 0.024 \\
 731 & 01 57 12.17 & +37 56 04.7 & 11.956 & 0.000 & 0.595 & 0.000 & 11.608 & 0.013 & 10.722 & 0.019 & 10.426 & 0.019 & 10.335 & 0.020 \\
 745 & 01 57 14.27 & +37 46 51.0 &  9.874 & 0.011 & 0.400 & 0.007 &  9.659 & 0.016 &  9.032 & 0.021 &  8.878 & 9.995 &  8.790 & 0.018 \\
 756 & 01 57 17.11 & +37 26 08.8 & 10.207 & 0.012 & 0.428 & 0.004 &  9.971 & 0.013 &  9.302 & 0.018 &  9.122 & 0.017 &  9.047 & 0.020 \\
 768 & 01 57 19.42 & +37 59 23.5 & 12.072 & 0.011 & 0.490 & 0.000 & 11.811 & 0.014 & 11.080 & 0.020 & 10.874 & 0.018 & 10.830 & 0.019 \\
 772 & 01 57 20.73 & +37 51 43.1 & 10.188 & 0.009 & 0.476 & 0.003 &  9.923 & 0.012 &  9.213 & 0.019 &  9.022 & 0.019 &  8.937 & 0.018 \\
 783 & 01 57 22.29 & +37 36 23.2 & 12.240 &       & 0.610 &       & 11.946 & 0.015 & 11.054 & 0.018 & 10.771 & 0.019 & 10.717 & 0.024 \\
 786 & 01 57 22.97 & +37 38 21.8 & 13.170 &       & 0.730 &       & 12.746 & 0.016 & 11.703 & 0.018 & 11.327 & 0.017 & 11.261 & 0.020 \\
 790 & 01 57 23.81 & +37 52 11.9 & 12.267 & 0.012 & 0.529 & 0.018 & 11.938 & 0.015 & 11.117 & 0.018 & 10.860 & 0.019 & 10.756 & 0.018 \\
\end{tabular}  
}  
\end{table*}

\begin{table*}
{\scriptsize
\begin{tabular}{rcccccccccccccc}
\hline
 ID & $\alpha$ & $\delta$ & \multicolumn{2} {c} {V} & \multicolumn{2} {c} {B-V}& \multicolumn{2} {c} {R} & \multicolumn{2} {c} {J}& \multicolumn{2} {c} {H}& \multicolumn{2} {c} {$K_{s}$}\\
\hline
 791 & 01 57 24.01 & +38 06 10.4 & 12.684 & 0.017 & 0.543 & 0.017 & 12.370 & 0.015 & 11.566 & 0.020 & 11.306 & 0.019 & 11.261 & 0.021 \\
 798 & 01 57 26.01 & +37 43 19.7 & 10.454 & 0.005 & 0.419 & 0.004 & 10.226 & 0.013 &  9.565 & 0.018 &  9.401 & 0.019 &  9.322 & 0.020 \\
 799 & 01 57 26.17 & +37 39 20.3 & 11.304 & 0.004 & 0.423 & 0.004 & 11.073 & 0.013 & 10.419 & 0.018 & 10.225 & 0.015 & 10.184 & 0.020 \\
 806 & 01 57 27.47 & +37 35 10.4 & 10.756 & 0.000 & 0.385 & 0.000 & 10.555 & 0.015 &  9.948 & 0.018 &  9.817 & 0.017 &  9.748 & 0.018 \\
 814 & 01 57 28.26 & +37 24 02.6 & 10.219 & 0.011 & 0.367 & 0.007 & 10.015 & 0.014 &  9.427 & 0.021 &  9.287 & 0.019 &  9.209 & 0.020 \\
 823 & 01 57 30.93 & +37 54 57.9 & 10.273 & 0.003 & 0.417 & 0.002 & 10.037 & 0.014 &  9.379 & 0.018 &  9.225 & 0.019 &  9.165 & 0.020 \\
 824 & 01 57 31.86 & +37 53 40.6 & 11.629 & 0.011 & 0.440 & 0.005 & 11.375 & 0.015 & 10.720 & 0.019 & 10.533 & 0.019 & 10.452 & 0.018 \\
 828 & 01 57 32.58 & +37 42 05.8 & 13.943 & 0.000 & 0.681 & 0.000 & 13.537 & 0.024 & 12.452 & 0.031 & 12.095 & 0.035 & 12.051 & 0.020 \\
 847 & 01 57 35.91 & +37 58 23.1 & 14.200 &       & 1.030 &       & 13.704 & 0.025 & 12.385 & 0.021 & 11.891 & 0.023 & 11.769 & 0.018 \\
 849 & 01 57 36.23 & +37 45 10.0 &  9.917 & 0.010 & 0.424 & 0.005 &  9.682 & 0.013 &  9.049 & 0.024 &  8.860 & 0.023 &  8.773 & 0.019 \\
 857 & 01 57 37.69 & +37 49 00.7 & 10.028 & 0.011 & 0.470 & 0.002 &  9.780 & 0.013 &  9.072 & 0.021 &  8.841 & 0.020 &  8.774 & 0.019 \\
 858 & 01 57 37.62 & +37 39 37.8 &  8.958 & 0.003 & 1.085 & 0.004 &  8.416 & 0.030 &  7.070 & 0.021 &  6.555 & 0.018 &  6.412 & 0.018 \\
 859 & 01 57 37.77 & +37 49 50.4 & 13.200 &       & 0.670 &       & 12.842 & 0.016 & 11.735 & 0.021 & 11.364 & 0.023 & 11.244 & 0.018 \\
 864 & 01 57 38.78 & +38 08 30.3 & 12.885 & 0.013 & 0.578 & 0.017 & 12.553 & 0.017 & 11.711 & 0.019 & 11.424 & 0.019 & 11.374 & 0.016 \\
 867 & 01 57 38.97 & +37 46 12.2 &  9.041 & 0.007 & 1.004 & 0.006 &  8.556 & 0.019 &  7.263 & 0.020 &  6.813 & 0.024 &  6.669 & 0.031 \\
 868 & 01 57 39.46 & +37 52 25.8 & 10.465 & 0.014 & 0.381 & 0.005 & 10.262 & 0.014 &  9.676 & 0.021 &  9.485 & 0.021 &  9.418 & 0.017 \\
 888 & 01 57 43.97 & +37 51 42.1 & 10.447 & 0.007 & 0.427 & 0.003 & 10.208 & 0.012 &  9.557 & 0.021 &  9.371 & 0.021 &  9.279 & 0.017 \\
 889 & 01 57 44.46 & +38 11 06.8 & 12.802 & 0.015 & 0.551 & 0.018 & 12.475 & 0.014 & 11.654 & 0.018 & 11.395 & 0.019 & 11.323 & 0.016 \\
 890 & 01 57 44.74 & +37 59 18.4 & 10.087 & 0.011 & 0.453 & 0.006 &  9.837 & 0.014 &  9.172 & 0.021 &  8.985 & 0.021 &  8.901 & 0.018 \\
 897 & 01 57 46.07 & +38 04 28.4 & 10.506 & 0.037 & 0.407 & 0.003 & 10.382 & 0.170 &  9.653 & 0.018 &  9.481 & 0.019 &  9.407 & 0.014 \\
 901 & 01 57 47.15 & +37 47 30.3 & 10.981 & 0.006 & 0.388 & 0.007 & 10.763 & 0.014 & 10.126 & 0.021 &  9.985 & 0.019 &  9.897 & 0.018 \\
 917 & 01 57 51.42 & +37 39 52.4 & 14.310 &       & 0.520 &       & 13.590 & 0.026 & 12.596 & 0.019 & 12.203 & 0.019 & 12.089 & 0.020 \\
 921 & 01 57 52.00 & +37 27 46.0 & 12.644 & 0.011 & 0.553 & 0.007 & 12.349 & 0.013 & 11.507 & 0.018 & 11.275 & 0.017 & 11.226 & 0.021 \\
 935 & 01 57 55.20 & +37 52 46.0 & 11.620 &       & 0.450 &       & 11.417 & 0.014 & 10.750 & 0.021 & 10.553 & 0.021 & 10.520 & 0.018 \\
 937 & 01 57 54.97 & +37 20 26.6 & 10.980 &       & 0.380 &       & 10.762 & 0.014 & 10.161 & 0.018 & 10.015 & 0.017 &  9.971 & 0.019 \\
 941 & 01 57 56.45 & +37 50 01.0 & 10.706 & 0.006 & 0.424 & 0.001 & 10.462 & 0.014 &  9.807 & 0.021 &  9.640 & 0.023 &  9.556 & 0.019 \\
 950 & 01 57 57.79 & +37 48 22.3 & 11.467 & 0.020 & 0.480 & 0.007 & 11.160 & 0.077 & 10.393 & 0.020 & 10.225 & 0.021 & 10.163 & 0.019 \\
 952 & 01 57 58.24 & +37 26 06.4 & 12.650 &       & 0.540 &       & 12.356 & 0.014 & 11.556 & 0.019 & 11.282 & 0.019 & 11.268 & 0.019 \\
 953 & 01 57 58.85 & +37 41 26.8 & 12.352 & 0.038 & 0.600 & 0.008 & 12.012 & 0.014 & 11.144 & 0.018 & 10.870 & 0.019 & 10.797 & 0.020 \\
 955 & 01 57 59.37 & +37 54 53.8 &  9.968 & 0.010 & 0.445 & 0.001 &  9.730 & 0.013 &  9.084 & 0.021 &  8.885 & 0.023 &  8.804 & 0.018 \\
 964 & 01 58 02.79 & +38 02 30.4 & 12.912 & 0.016 & 0.582 & 0.014 & 12.568 & 0.015 & 11.668 & 0.019 & 11.385 & 0.019 & 11.273 & 0.018 \\
 983 & 01 58 06.31 & +37 38 06.6 & 13.110 &       & 0.610 &       & 12.709 & 0.017 & 11.756 & 0.019 & 11.401 & 0.019 & 11.314 & 0.020 \\
 988 & 01 58 07.71 & +37 39 57.0 & 10.934 & 0.008 & 0.373 & 0.004 & 10.714 & 0.016 & 10.144 & 0.019 &  9.987 & 0.019 &  9.953 & 0.017 \\
 993 & 01 58 09.26 & +37 28 35.5 & 13.590 & 0.000 & 0.708 & 0.000 & 13.226 & 0.018 & 12.271 & 0.021 & 11.923 & 0.021 & 11.855 & 0.021 \\
 999 & 01 58 10.66 & +37 24 05.9 & 13.550 &       & 0.670 &       & 13.189 & 0.019 & 12.236 & 0.021 & 11.899 & 0.023 & 11.819 & 0.019 \\
1000 & 01 58 11.43 & +37 39 33.4 & 11.410 & 0.012 & 0.419 & 0.006 & 11.175 & 0.014 & 10.568 & 0.021 & 10.379 & 0.019 & 10.332 & 0.018 \\
1003 & 01 58 12.27 & +37 32 38.2 & 11.193 & 0.003 & 0.472 & 0.003 & 10.933 & 0.014 & 10.251 & 0.021 & 10.013 & 0.018 &  9.942 & 0.017 \\
1007 & 01 58 13.40 & +38 11 41.5 & 13.018 & 0.025 & 0.556 & 0.042 & 12.641 & 0.016 & 11.742 & 0.020 & 11.467 & 0.017 & 11.375 & 0.016 \\
1008 & 01 58 12.71 & +37 34 40.4 & 10.959 & 0.005 & 0.371 & 0.004 & 10.743 & 0.015 & 10.188 & 0.021 & 10.009 & 0.018 &  9.964 & 0.019 \\
1012 & 01 58 12.94 & +37 15 20.2 & 12.417 & 0.015 & 0.539 & 0.000 & 12.113 & 0.016 & 11.336 & 0.020 & 11.083 & 0.023 & 11.047 & 0.019 \\
1017 & 01 58 15.33 & +37 33 19.6 & 13.260 &       & 0.660 &       & 12.920 & 0.017 & 12.014 & 0.021 & 11.673 & 0.019 & 11.602 & 0.019 \\
1023 & 01 58 16.88 & +37 38 15.9 & 11.250 & 0.015 & 0.424 & 0.010 & 11.005 & 0.013 & 10.407 & 0.021 & 10.203 & 0.019 & 10.159 & 0.019 \\
1026 & 01 58 19.00 & +38 32 14.0 & 11.089 & 0.016 & 0.377 & 0.017 & 10.851 & 0.010 & 10.303 & 0.020 & 10.134 & 0.023 & 10.079 & 0.017 \\
1027 & 01 58 18.42 & +38 06 54.0 & 12.597 & 0.020 & 0.541 & 0.036 & 12.282 & 0.022 & 11.497 & 0.020 & 11.237 & 0.017 & 11.184 & 0.018 \\
1083 & 01 58 27.61 & +37 35 22.2 & 11.920 & 0.011 & 0.484 & 0.007 & 11.676 & 0.014 & 10.977 & 0.021 & 10.721 & 0.019 & 10.656 & 0.018 \\
1089 & 01 58 29.84 & +37 51 37.4 &  9.303 & 0.007 & 0.963 & 0.008 &  8.836 & 0.021 &  7.621 & 0.026 &  7.166 & 0.026 &  7.040 & 0.021 \\
1107 & 01 58 34.42 & +37 40 15.1 & 13.660 &       & 0.660 &       & 13.273 & 0.020 & 12.299 & 0.022 & 11.923 & 0.019 & 11.884 & 0.021 \\
1117 & 01 58 36.91 & +37 45 10.6 &  9.598 & 0.002 & 0.406 & 0.002 &  9.370 & 0.012 &  8.747 & 0.027 &  8.559 & 0.026 &  8.502 & 0.016 \\
1123 & 01 58 38.12 & +37 32 15.7 & 11.507 & 0.000 & 0.417 & 0.007 & 11.278 & 0.014 & 10.647 & 0.020 & 10.467 & 0.019 & 10.395 & 0.019 \\
1129 & 01 58 40.07 & +37 38 05.1 & 11.910 & 0.011 & 0.481 & 0.007 & 11.629 & 0.014 & 10.911 & 0.021 & 10.691 & 0.019 & 10.599 & 0.018 \\
1151 & 01 58 47.96 & +38 26 08.2 & 10.063 & 0.013 & 0.444 & 0.014 &  9.788 & 0.014 &  9.158 & 0.021 &  8.947 & 0.018 &  8.912 & 0.019 \\
1161 & 01 58 50.00 & +37 59 46.6 & 14.600 &       & 0.680 &       & 14.060 & 0.040 & 12.715 & 0.019 & 12.215 & 0.019 & 12.108 & 0.020 \\
1165 & 01 58 50.44 & +37 20 52.0 & 10.462 & 0.000 & 0.390 & 0.000 & 10.239 & 0.013 &  9.643 & 0.020 &  9.485 & 0.023 &  9.436 & 0.020 \\
1172 & 01 58 52.93 & +37 48 57.0 &  9.060 & 0.003 & 1.036 & 0.006 &  8.529 & 0.036 &  7.270 & 0.019 &  6.795 & 0.023 &  6.643 & 0.018 \\
1178 & 01 58 53.94 & +37 34 42.6 & 13.390 &       & 0.830 &       & 13.028 & 0.015 & 11.711 & 0.021 & 11.268 & 0.019 & 11.197 & 0.022 \\
1196 & 01 58 57.32 & +37 39 40.9 & 13.810 &       & 0.910 &       & 13.339 & 0.019 & 12.150 & 0.021 & 11.703 & 0.017 & 11.611 & 0.022 \\
1204 & 01 58 59.87 & +38 01 18.9 & 11.680 &       & 0.460 &       & 11.446 & 0.014 & 10.796 & 0.021 & 10.632 & 0.023 & 10.577 & 0.019 \\
1263 & 01 59 14.82 & +38 00 55.2 &  9.006 & 0.011 & 1.019 & 0.007 &  8.483 & 0.025 &  7.199 & 0.019 &  6.767 & 0.018 &  6.648 & 0.016 \\
1270 & 01 59 18.04 & +37 49 49.4 & 14.090 &       & 0.700 &       & 13.457 & 0.060 & 12.552 & 0.019 & 12.159 & 0.021 & 12.069 & 0.022 \\
1284 & 01 59 19.91 & +37 23 23.1 & 12.893 & 0.000 & 0.677 & 0.000 & 12.476 & 0.013 & 11.472 & 0.021 & 11.140 & 0.021 & 11.052 & 0.020 \\
1296 & 01 59 26.08 & +37 40 39.9 & 14.750 &       & 0.530 &       & 14.111 & 0.035 & 13.123 & 0.023 & 12.767 & 0.022 & 12.676 & 0.027 \\
1304 & 01 59 29.63 & +38 16 04.3 & 11.341 & 0.008 & 0.412 & 0.011 & 11.076 & 0.014 & 10.485 & 0.022 & 10.293 & 0.022 & 10.251 & 0.020 \\
1365 & 01 59 47.27 & +37 49 53.8 & 13.290 &       & 0.710 &       & 12.935 & 0.018 & 12.005 & 0.022 & 11.696 & 0.021 & 11.669 & 0.019 \\
1407 & 01 59 56.83 & +37 58 10.5 & 12.949 & 0.000 & 0.552 & 0.000 & 12.630 & 0.016 & 11.782 & 0.027 & 11.575 & 0.034 & 11.454 & 0.023 \\
1474 & 02 00 21.98 & +38 02 41.0 & 10.698 & 0.009 & 0.355 & 0.016 & 10.441 & 0.013 &  9.894 & 0.027 &  9.730 & 0.030 &  9.692 & 0.022 \\
1602 & 02 01 05.97 & +37 42 23.6 &  9.961 & 0.011 & 0.462 & 0.014 &  9.671 & 0.013 &  8.978 & 0.020 &  8.786 & 0.032 &  8.733 & 0.023 \\
\hline
\end{tabular}  
}  
\end{table*}

\end{document}